\documentclass[twocolumn,superscriptaddress,aps,preprintnumbers,amsmath,amssymb,sort&compress,nofootinbib
]{revtex4-1}

\usepackage{amsfonts,amsmath,amssymb,ascmac,bm,tensor}
\usepackage{comment}
\usepackage{ifpdf}
\usepackage{slashed}
\usepackage{color}
\usepackage[mathscr]{eucal}
\usepackage[utf8]{inputenc}

\usepackage{cancel}

\ifpdf
  \usepackage{graphicx}     
  \usepackage[bookmarksopen,colorlinks=true,linkcolor=bblue,citecolor=bblue,urlcolor=ppink]{hyperref}
\else     
  \usepackage[dvipdfmx]{graphicx}     
  \usepackage[dvipdfmx,bookmarksopen,colorlinks=true,linkcolor=bblue,citecolor=ppink,urlcolor=darkred]{hyperref}
\fi

\definecolor{red}{rgb}{1,0,0}
\definecolor{darkred}{rgb}{0.6,0,0}
\definecolor{darkgreen}{rgb}{0.992447,0.623778,0.034597}
\definecolor{ppink}{rgb}{1,0.4,0.4}
\definecolor{bblue}{rgb}{0.284602,0.317763,0.963947}

\newcommand{\1}{\mbox{1}\hspace{-0.25em}\mbox{l}}

\newcommand{\dd}{\mathrm{d}}

\makeatletter
\newcommand\footnoteref[1]{\protected@xdef\@thefnmark{\ref{#1}}\@footnotemark}
\makeatother

\allowdisplaybreaks[1]

\begin{document}


\title{
Gravitational waves induced by scalar perturbations \\as probes of the small-scale primordial spectrum
}
\author{Keisuke Inomata}
\affiliation{ICRR, University of Tokyo, Kashiwa, 277-8582, Japan}
\affiliation{Kavli IPMU (WPI), UTIAS, University of Tokyo, Kashiwa, 277-8583, Japan}
\author{Tomohiro Nakama}
\affiliation{Institute for Advanced Study, The Hong Kong University of Science and Technology, \\Clear Water Bay, Kowloon, Hong Kong}

\begin{abstract}
\noindent
Compared to primordial perturbations on large scales, roughly larger than $1$ megaparsec, those on smaller scales are not severely constrained.
We revisit the issue of probing small-scale primordial perturbations using gravitational waves (GWs), based on the fact that, 
when large-amplitude primordial perturbations on small scales exist, GWs with relatively large amplitudes are induced at second order in scalar perturbations, and these induced GWs can be probed by both existing and planned gravitational-wave projects.
We use accurate methods to calculate these induced GWs and take into account sensitivities of different experiments to induced GWs carefully, to report existing and expected limits on the small-scale primordial spectrum.
\end{abstract}

\date{\today}
\maketitle
\preprint{IPMU 18-0200}

\section{Introduction}
\label{sec:intro}

In inflationary cosmology, the Universe experienced an early stage of accelerated expansion, 
during which primordial perturbations with a vast range of wavelengths are produced from quantum fluctuations. Thanks to the recent observations, we have determined the cosmological parameters including those characterizing these primordial fluctuations well and entered into an era of precision cosmology.
On large scales with comoving wavenumbers $k \lesssim 1$Mpc$^{-1}$, the amplitude of the curvature perturbation, which describes one kind of primordial perturbations,\footnote{
Isocurvature perturbation is the other kind of primordial perturbations, which is tightly constrained on large scales by observations~\cite{Akrami:2018odb}.
} has been precisely determined by observations of the cosmic microwave background (CMB) and large-scale structure (LSS) of the Universe as $\mathcal P_\mathcal{R} \simeq 2.1\times 10^{-9}$ at $k=0.05\, \mathrm{Mpc}^{-1}$ with a slight scale dependence~\cite{Aghanim:2018eyx}.

On the other hand, it is difficult to determine the small-scale ($k \gtrsim 1$Mpc$^{-1}$) curvature perturbations with CMB or LSS due to the limited sensitivity and resolution of the experiments, as well as the Silk damping or highly non-linear, complicated evolutions of inhomogeneities, as a result of which the information about primordial fluctuations is partially or entirely lost.
Hence, the constraints on the small-scale perturbations from these observations are significantly weaker or virtually non-existent than those on the large-scale ones.

The properties of primordial fluctuations, including those on small scales which have not been well investigated as mentioned above, depend on inflation mechanisms \cite{Kawasaki:1998vx,Kasuya:2009up,Garcia-Bellido:2017mdw}.
Thus, the study of the small-scale perturbations can shed light on the nature of the inflation.
In addition, large-amplitude perturbations on small scales could lead to unique compact objects, such as primordial black holes (PBHs) (see Ref.~\cite{Sasaki:2018dmp} for a recent review) 
 and ultracompact minihalos (UCMHs)~\cite{Ricotti:2009bs,Scott:2009tu,Kohri:2014lza,Nakama:2017qac}.
In particular, PBHs have recently been attracting a lot of attention because PBHs are one of the candidates for dark matter (DM)~\cite{Inomata:2017okj,Ballesteros:2017fsr,Inomata:2017vxo} and also for the black holes detected by the direct observations of gravitational waves (GWs)~\cite{Bird:2016dcv,Clesse:2016vqa,Sasaki:2016jop,Eroshenko:2016hmn}.
Therefore it is increasingly important to discuss how large perturbations can be on small scales also from the viewpoint of the nature of the dark matter or gravitational-wave astrophysics.

The small-scale perturbations have been constrained as follows;
non-detection of CMB distortions, $\mathcal P_\mathcal{R} \lesssim 10^{-4}$ on $k\lesssim10^4$Mpc$^{-1}$~\cite{Chluba:2012gq,Chluba:2012we,Kohri:2014lza};  
consistency with the observed abundance of the light elements produced at big bang nucleosynthesis (BBN), $\mathcal P_\mathcal{R} \lesssim 10^{-2}$ on $10^4$Mpc$^{-1} \lesssim k \lesssim 10^5$Mpc$^{-1}$~\cite{Jeong:2014gna,Nakama:2014vla,Inomata:2016uip}; 
non-detection of gamma rays from UCMHs, $\mathcal P_\mathcal{R} \lesssim 10^{-6}$ on $k \lesssim 10^7$Mpc$^{-1}$~\cite{Bringmann:2011ut}; 
and non-detection of PBHs, $\mathcal P_\mathcal{R} \lesssim 10^{-2}$ over a wide range of scales~\cite{Josan:2009qn}.
However, some of these constraints are not only weak but also uncertain compared to those on large scales.
For example, the constraints from UCMHs strongly depend on the properties of DM~\cite{Bringmann:2011ut,Nakama:2017qac} and their profiles~\cite{Gosenca:2017ybi,Delos:2017thv}, which have not been fully understood, 
and the constraints from PBHs are based on some simplifying assumptions about the relation between the PBH abundance and the amplitude of curvature perturbations, hence they also involve some uncertainties~\cite{Nakama:2013ica,Ando:2018qdb,Yoo:2018kvb}.

In this paper, we focus on probing small-scale primordial fluctuations by GWs induced at second order in curvature perturbations (see \cite{Assadullahi:2009jc} and references therein), noting that, 
although the evolutions of GWs are independent of curvature perturbations at the linear order, they depend on curvature perturbations at second order.
These GWs induced at second order in curvature perturbations can be constrained by the current and future observations, 
such as pulsar timing array (PTA) observations (EPTA~\cite{Lentati:2015qwp}, PPTA~\cite{Shannon:2015ect}, NANOGrav~\cite{Arzoumanian:2015liz}, SKA~\cite{Moore:2014lga,Janssen:2014dka}), 
second-generation GW interferometers (advanced LIGO (aLIGO)~\cite{Abbott:2017xzg}, Virgo~\cite{virgo},  KAGRA~\cite{kagra}), 
space-based GW interferometers (LISA~\cite{Sathyaprakash:2009xs,Moore:2014lga,Audley:2017drz}, DECIGO~\cite{Seto:2001qf,Yagi:2011wg}, BBO~\cite{phinney2003big,Yagi:2011wg}), and third-generation GW interferometers (Einstein Telescope~\cite{Sathyaprakash:2009xs,Moore:2014lga,ET_sense}, Cosmic Explorer~\cite{Evans:2016mbw}).
Then using the limits on the induced GWs, we can obtain limits on the curvature perturbations on small scales.

This topic was first discussed in a pioneering work by Assadullahi \& Wands~\cite{Assadullahi:2009jc}.\footnote{
GWs induced at second order in scalar perturbations are also discussed in cases where a cosmologically interesting amount of PBHs is produced~\cite{Saito:2008jc, Saito:2009jt, Alabidi:2012ex,Alabidi:2013wtp,Garcia-Bellido:2016dkw}. 
}
We note, however, that the formulation to calculate induced GWs has been updated since then. In particular, sophisticated analytical formulae to calculate induced GWs have recently been derived by Kohri \& Terada~\cite{Kohri:2018awv} (see also Ref.~\cite{Espinosa:2018eve}).
The induced GWs predicted by their formula differ from those predicted by the relation used in Ref.~\cite{Assadullahi:2009jc} by an order of magnitude for the same curvature perturbations as discussed later. Since the study of small-scale primordial perturbations is  one of the recent hot topics in cosmology and the projects for GW observations are expected to make more and more progress in the near future, 
it is worth reconsidering the limits using the recent analytical formula and also using updated, expected sensitivities of planned GW detectors of different kinds. We also report new limits on the small-scale primordial power spectrum, obtained from the null detection of stochastic GWs by recent gravitational-wave experiments. 

\section{Formalism for induced GWs}
\label{sec:formalism}

In this section, we briefly review the equations to calculate GWs induced at second order in curvature perturbations (see also Refs.~\cite{Kohri:2018awv,Inomata:2016rbd}).
Throughout this paper, we assume that the GWs are induced during the radiation-dominated (RD) era.
Since the induced GWs can be enhanced due to an early matter dominated era, as a result of non-decaying subhorizon perturbations (see Ref.~\cite{Kohri:2018awv} and references therein), this assumption leads to conservative limits.\footnote{
The comoving scales reentering the horizon during the late-time matter-dominated era are large enough ($k < k_\text{eq} = 0.0103\, \text{Mpc}^{-1}$~\cite{Aghanim:2018eyx}) to be probed by CMB anisotropy observations well.
On the other hand, the scales potentially affected by a possible early matter-dominated era, preceding the RD era, could overlap the scales we consider in this paper ($\mathcal O(10)$Mpc$^{-1} < k < \mathcal O(10^{20})$Mpc$^{-1}$), 
if the reheating temperature is less than roughly $10^{13}$GeV.}
We also assume that the curvature perturbations follow the Gaussian distribution\footnote{ The GWs induced by scalar perturbations with non-Gaussianities are discussed in Ref.~\cite{Nakama:2016gzw,Garcia-Bellido:2017aan,Cai:2018dig,Unal:2018yaa}.}
and we take the conformal Newtonian gauge in this work.\footnote{
The gauge dependence of the induced GWs is discussed in Ref.~\cite{Hwang:2017oxa}, for those induced during the late-time matter domination era.
 We expect, however, that GWs induced during the RD era, induced mostly at horizon reentry, would not change so significantly by the choice of the gauge, though careful calculations to clarify this issue would be merited. 
 The calculations of Ref.~\cite{Hwang:2017oxa} show that the gauge dependence is not so significant for modes of GWs induced during the late-time matter domination whose wavelengths are comparable to the horizon at each moment in time. 
This implies that the gauge dependence of the subhorizon evolution of scalar perturbations is probably the primary cause of the gauge dependence of the GWs induced during the 
late-time matter domination found in their work.
Since, unlike during the matter-dominated era, the scalar perturbations decay on subhorizon scales during the RD era, we expect the gauge dependence of the GWs induced during RD era would not be so significant.
}
The energy density of GWs per logarithmic interval of $k$ is
\begin{align}
\Omega_{\rm{GW}} (\eta, k) &= \frac{\rho_{\rm{GW}} (\eta, k) } { \rho_{\rm{tot}}(\eta)} \nonumber\\ 
&=
\frac{1}{24} \left( \frac{k}{a(\eta) H(\eta) } \right)^2 
\overline{\mathcal P_{h} (\eta,k)},
\label{eq:gw_formula}
\end{align}
where the overline indicates time average and  $\mathcal P_{h}$ represents the power spectrum of induced GWs given by
\begin{align}
\overline{\mathcal P_{h} (\eta,k)} \simeq 4\int^\infty_0 \dd v &\int^{1+v}_{|1-v|} \dd u \left( \frac{4v^2 - (1+v^2 - u^2)^2}{4vu}  \right)^2 \nonumber \\
& \times \overline{I^2 (v,u,k\eta)}\mathcal P_\zeta(k v) \mathcal P_\zeta(ku).
\label{eq:p_h_formula}
\end{align}
Changing the variables to $t=u+v-1$ and $s=u-v$, we can rewrite Eq.~(\ref{eq:p_h_formula}) as
\begin{align}
\overline{\mathcal P_{h} (\eta,k)} \simeq 2\int^\infty_0 \dd t &\int^1_{-1} \dd s \left( \frac{t(2+t)(s^2 -1)}{(1-s+t)(1+s+t)}  \right)^2 \nonumber \\
& \times \overline{I^2 (v,u,k\eta)}\mathcal P_\zeta(k v) \mathcal P_\zeta(ku).
\label{eq:p_h_formula2}
\end{align}
The function $\overline{I^2}$ in the subhorizon limit ($x=k\eta\rightarrow \infty$) is
\begin{widetext}
\begin{align}
 \overline{I^2 (v,u, x \rightarrow \infty)} =& \frac{1}{2} \left( \frac{3(u^2+v^2 -3)}{4u^3 v^3 x} \right)^2 \left( \left( -4uv + (u^2 +v^2 -3)\, \text{log} \left| \frac{3 - (u+v)^2 }{3-(u-v)^2} \right| \right)^2 \nonumber \right.\\
 & \qquad \qquad \qquad \qquad \qquad  \left. + \pi^2 (u^2 + v^2 -3 )^2 \Theta( v+u- \sqrt{3}) \right) \\
 =& \frac{288 (-5 + s^2 +t(2+t))^2}{x^2 (1-s+t)^6 (1+s+t)^6} \left( \frac{\pi^2}{4}  (-5 + s^2 +t(2+t))^2 \Theta(t-(\sqrt{3}-1)) \right. \nonumber \\
 & \left. + \left( -(t-s+1)(t+s+1) + \frac{1}{2}  (-5 + s^2 +t(2+t))\, \text{log} \left| \frac{-2+t(2+t)}{3-s^2} \right| \right)^2 \right),
 \label{eq:I_approx}
 \end{align}
\end{widetext}
where $\Theta$ denotes the Heaviside step function. We have confirmed that induced GWs calculated from these approximations indeed coincide well with those obtained from numerical integrations of the exact integrand, 
using Eqs.~(A44) and (A45) of Ref.~\cite{Inomata:2016rbd},
with the time average taken after the integrations.

During the RD era, induced GWs are produced mainly around horizon reentry, without growing any more after that because the gravitational potential decays after the horizon reentry. This can be seen from the above formula, by noting that the factor $x^2$ in the denominator is in the end canceled by the factor $(k/aH)^2$ in Eq.~(\ref{eq:gw_formula}), given the relation $a(\eta)H(\eta) = \eta^{-1}$, which holds during the RD era.
This indicates that the GWs, being time independent, are no longer induced on subhorizon scales. 
We define $\eta_c$ as the moment when $\Omega_{\mathrm{GW}}$ stops growing, which is shortly after the horizon reentry,  
and note that $\eta_c$ is earlier than the beginning of the late-time matter domination, since we consider only those modes which reenter the horizon well before then.

Taking into account the evolutions of $\Omega_\text{GW}$ after the matter-radiation equality and the change in relativistic degrees of freedom, we can derive the relation between the density parameter at $\eta_0$ (today) and that at $\eta_c$ as~\cite{Ando:2018qdb}
\begin{align}
\Omega_{\rm{GW}} (\eta_0, k)  = 0.83 \left( \frac{g_c}{10.75} \right)^{-1/3} \Omega_{r,0} \Omega_{\rm{GW}}(\eta_c, k),\label{latetime}
\end{align}
where $\Omega_{r,0}$ is the current energy density parameter of radiation, $g$ is the effective degrees of freedom contributing to the total radiation energy density and the subscript ``$c$'' indicates the value at $\eta_c$. In order to obtain $g_c$, normally given as a function of the temperature $T$ (see e.g. Ref. \cite{Aghanim:2018eyx}), for each wavenumber, we need the relation between the scale $k$ that enters the horizon at $\eta$ and the temperature at that time. As shown in App.~\ref{app:relation_k_t}, it is
\begin{align}
\frac{k}{k_\text{eq}} = 2(\sqrt{2}-1) \left(\frac{g_\text{s,eq}}{g_\text{s}}\right)^{1/3} \left( \frac{g}{g_\text{eq}} \right)^{1/2} \frac{T}{T_\text{eq}},
\label{eq:relation_k_t}
\end{align}
where $g_s$ denotes the effective degrees of freedom for the entropy density and the subscript ``eq'' means the value at the matter-radiation equality time.
We take $k_\text{eq} = 0.0103$\,Mpc$^{-1}$, $T_\text{eq} = 8.0\times10^{-7}$\,MeV,\footnote{
$T_\text{eq}$ is given by $T_\text{eq} = (1+z_\text{eq}) 2.725$\,K, where $z_\text{eq} \simeq 2.4 \times 10^4\, \Omega_\text{m} h^2 (\simeq 3409)$ \cite{Aghanim:2018eyx,Dodelson:1282338}.
} $g_\text{s,eq} = 3.91$ and $g_\text{eq} = 3.38$~\cite{Aghanim:2018eyx,Kolb:206230,Dodelson:1282338}.

\section{Constraints on induced GWs}

\begin{figure}
\centering
\includegraphics[width=0.45 \textwidth]{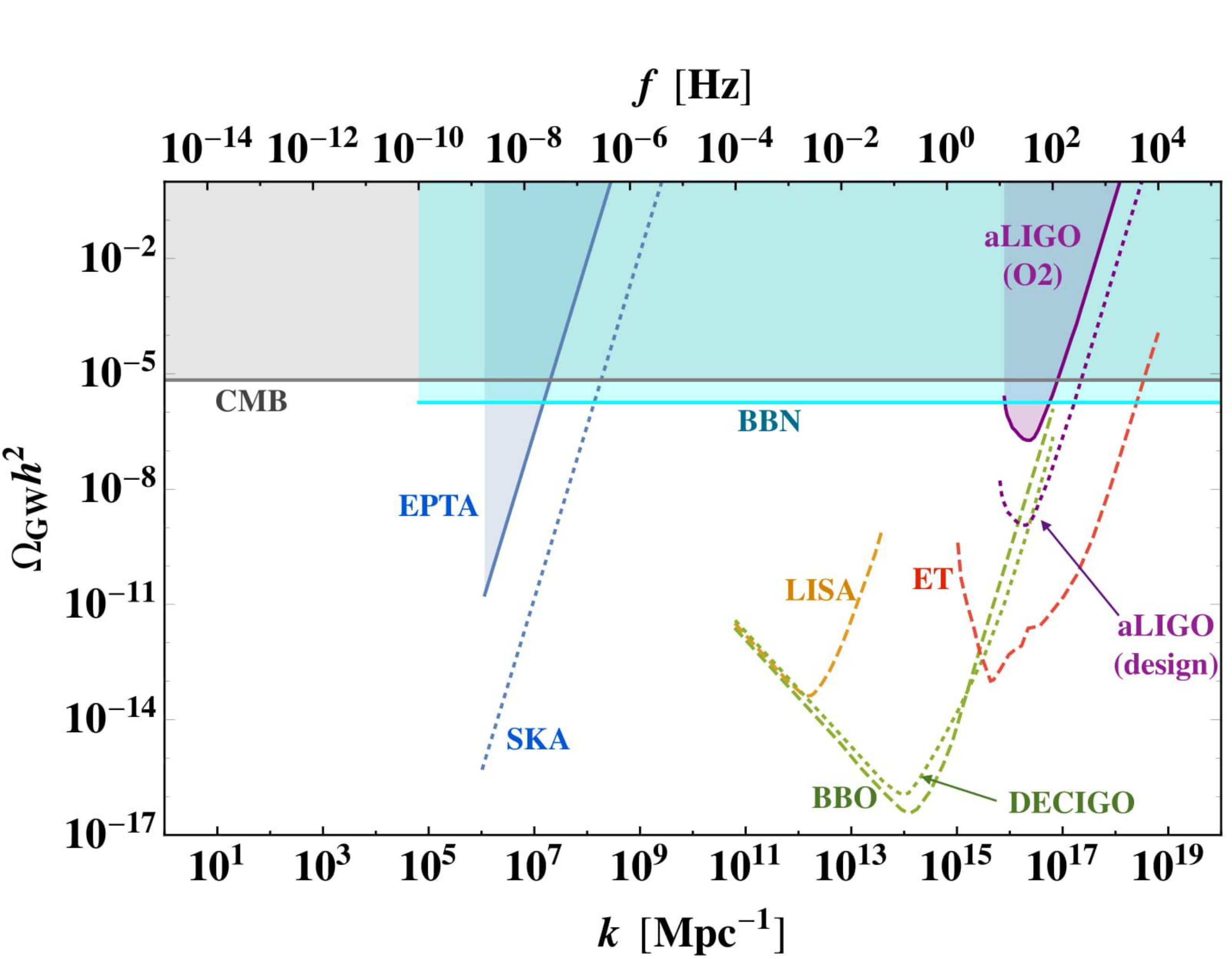}
\caption{
 	Effective sensitivities to stochastic GWs of current and future gravitational-wave projects and constraints on GWs.
	Note that, except for the plots labeled as CMB and BBN, we plot $\Omega_\text{GW,eff} h^2/\sqrt{T f/10}$ to illustrate the effective sensitivities, where $\Omega_\text{GW,eff}$ is defined as Eq.~(\ref{eq:eff_omegagw}) in the weak-signal limit.
	Hence these curves should be distinguished from the power-law integrated curves of Ref.~\cite{Thrane:2013oya}.
        We include 
         ongoing PTA observations (EPTA~\cite{Lentati:2015qwp}), 
	a future PTA observation by SKA~\cite{Moore:2014lga,Janssen:2014dka}, 
	a second-generation ground-based GW interferometer (advanced LIGO, for which both the limits from the O2 run and design sensitivity are shown~\cite{Aasi:2013wya,Thrane:2013oya,Abbott:2017xzg}), 
	 space-based GW interferometers (LISA~\cite{Thrane:2013oya,Moore:2014lga,Audley:2017drz}, BBO~\cite{phinney2003big,Yagi:2011wg,Thrane:2013oya}, DECIGO~\cite{Kuroyanagi:2014qza}),
	  and finally third-generation ground-based GW interferometers (Einstein Telescope (ET)~\cite{Sathyaprakash:2009xs,Moore:2014lga,ET_sense}). 
	 There are also other constraints from CMB and LSS~\cite{Smith:2006nka}, as well as BBN~\cite{Kohri:2018awv}, which should be noted to be existing limits on stochastic GWs.
	We take the observation time $T$ as 18 years for EPTA~\cite{Lentati:2015qwp}, 20 years for SKA, 4 months for aLIGO(O2)~\cite{Abbott:2017xzg}, and 1 year for the others.
	The shaded regions are already excluded by the existing observational data.
	See text for more details about each project.
}
\label{fig:gw_const_summary}
\end{figure}

In this section, we briefly review techniques to investigate stochastic GW backgrounds with observations before applying them to induced GWs. 
Our analysis is basically based on that in Ref.~\cite{Thrane:2013oya}. When multiple detectors or pulsars whose noise is uncorrelated are available, it is highly beneficial to do a cross-correlation to look for the correlated signal due to a stochastic GW background. 
In this section, we use frequencies of GWs instead of their wave numbers, which are related as $f= 1.546\times10^{-15} (k/1\text{Mpc}^{-1})$\,Hz.
First, the signal-to-noise ratio $\rho$ for a collection of detectors or pulsars labeled by $I$ and $J$ receiving stochastic GWs is~\cite{Allen:1997ad}
\begin{align}
\rho = \sqrt{2T} \left[ \int^{f_\text{max}}_{f_\text{min}} \dd f \, \sum^M_{I=1} \sum^M_{J>I} \frac{\Gamma^2_{IJ}(f) S_h^2(f)}{P_{nI}(f) P_{nJ}(f)} \right]^{1/2},
\label{eq:rho_def1}
\end{align}
where $M$ is the number of detectors or pulsars, $T$ is the observation time, $P_n$ is the noise power spectrum, $\Gamma_{IJ}$ is the overlap reduction function between $I$-th and $J$-th detectors or pulsars, $S_h$ is the power spectral density of GWs, and $f_\text{max}$ and $f_\text{min}$ are the maximum and minimum observation frequencies respectively.
Note that Eq.~(\ref{eq:rho_def1}) is valid only in the weak-signal limit \cite{Siemens:2013zla}, which may not be applied to PTA experiments, depending on projects.
We will discuss this issue later.
Here, we define the effective sensitivity curve for the GW energy density as
\begin{align}
\Omega_\text{GW,eff}(f) H_0^2 = \frac{2\pi^2}{3} f^3 \left[ \sum^M_{I=1} \sum^M_{J>I} \frac{\Gamma^2_{IJ}(f)}{P_{nI}(f) P_{nJ}(f)} \right]^{-1/2},
\label{eq:eff_omegagw}
\end{align}
where $H_0$ is the Hubble constant.
Then, we can rewrite Eq.~(\ref{eq:rho_def1}) as
\begin{align}
\rho = \sqrt{2T} \left[ \int^{f_\text{max}}_{f_\text{min}} \dd f \, \left(\frac{\Omega_\text{GW}(f)}{\Omega_\text{GW,eff}(f)}\right)^2 \right]^{1/2},
\label{eq:rho_def}
\end{align}
where we define $\Omega_\text{GW}(f)H_0^2 \equiv 2\pi^2 f^3 S_h(f)/3 $.
In Fig.~\ref{fig:gw_const_summary}, we plot $\Omega_\text{GW,eff} h^2/\sqrt{Tf/10}$ as the effective sensitivity curves for each project using GW interferometers or pulsars.\footnote{
There are also proposals for an atomic GW interferometric sensor probing GWs in the 50 mHz - 10 Hz \cite{Hogan:2010fz} and optically levitated sensors to detect high-frequency (50 Hz - 300 kHz) GWs \cite{Arvanitaki:2012cn}. } 
 In addition, we also plot constraints on $\Omega_\text{GW}h^2$ from CMB, LSS, and BBN in Fig.~\ref{fig:gw_const_summary}, which are different from $\Omega_\text{GW,eff} h^2$.
We ignore spikes in the effective sensitivity for simplicity, which arise from zeros of overlap reduction functions. Because of the frequency integration above, neglecting such spikes would not cause huge errors. 
Note that some of the effective sensitivity curves in Fig.~\ref{fig:gw_const_summary} are based on approximations, which we explain below.

{\bf Advanced LIGO.}  
$\Omega_\text{GW,eff}$ for the aLIGO design sensitivity is calculated in Ref.~\cite{Thrane:2013oya}, which is based on
the correlation between the two detectors in Hanford and Livingston.
We also consider the latest results of the O2 run.
We obtain an approximation of $\Omega_\text{GW,eff}$ for the O2 run by renormalizing the amplitude of and rescaling the frequency dependence of $\Omega_\text{GW,eff}$ for the design sensitivity 
so that the minimum of the power-law integrated curve~\cite{Thrane:2013oya} calculated from the approximation reproduces the minimum of the power-law integrated curve for the O2 run presented in Ref.~\cite{Abbott:2017xzg}.
If Virgo \cite{virgo} and KAGRA \cite{kagra} reach the same level of sensitivity as aLIGO in the future, we can make use of them taking cross-correlations between more detectors and the effective sensitivity would be stronger.

{\bf Space-based interferometers.} 
$\Omega_\text{GW,eff}$ for DECIGO is calculated in Refs.~\cite{Kudoh:2005as,Kuroyanagi:2014qza}, which is based on the correlation between the Michelson interferometers located at opposite vertices of the star-of-David form.
$\Omega_\text{GW,eff}$ for BBO is also calculated in Ref.~\cite{Thrane:2013oya}.\footnote{
Since BBO and DECIGO have similar sensitivity curves~\cite{Yagi:2011wg}, we extrapolate the sensitivity curve of BBO in Ref.~\cite{Thrane:2013oya} to cover the same range of frequency as that of DECIGO in Ref.~\cite{Kuroyanagi:2014qza}. 
}

For LISA, although the cross-correlation technique cannot be applied due to its configuration~\cite{Thrane:2013oya,Cutler:1997ta},
its configuration could make it possible to disentangle stochastic GWs and  instrumental noise~\cite{Tinto:2001ii,Hogan:2001jn,Adams:2010vc,Adams:2013qma}.
Assuming instrumental noise and/or astrophysical foreground are removed perfectly, to estimate the signal-to-ratio by Eq. (\ref{eq:rho_def}), we can redefine $\Omega_\text{GW,eff}$ for LISA as~\cite{Thrane:2013oya}
\begin{align}
\Omega_\text{GW,eff}(f) H_0^2 = \sqrt{2} \, \frac{2\pi^2}{3} f^3 \frac{P_n(f)}{\Gamma(f)},
\label{eq:eff_omegagw_lisa}
\end{align}
where $\Gamma(f)$ is the transfer function of the detector and $P_n(f)$ is the noise power spectrum.
Note that, since we assume a single detector instead of two, the factor $\sqrt{2}$ appears in Eq.~(\ref{eq:eff_omegagw_lisa})~\cite{Thrane:2013oya}. 
In this paper, to present crude estimations  of the constraints on curvature perturbations from LISA, we use $\Omega_\text{GW,eff}$ for LISA obtained in Ref.~\cite{Thrane:2013oya}, which is based on the above relation.

{\bf Third-generation ground-based interferometers.}
Einstein Telescope (ET) is proposed to have three detectors configured in a triangle similarly to LISA, each of which consists of two interferometers.
Therefore the noise removal techniques proposed for LISA could possibly be applied to ET.
We also use Eq.~(\ref{eq:eff_omegagw_lisa}) for ET to obtain $\Omega_\text{GW,eff}$, with the sensitivity curve given in Refs.~\cite{Sathyaprakash:2009xs,Moore:2014lga,ET_sense}.
On the other hand, since Cosmic Explorer (CE) is proposed to have L-shaped geometry, as aLIGO, we cannot use the noise removal techniques mentioned above for CE.
Therefore we do not consider CE in the following, but if we have more than one CE-like detector in the future, we would be able to use the cross-correlation technique and probe stochastic GWs with them~\cite{Regimbau:2016ike}.
In this case, the constraints on curvature perturbations from CE-like detectors would be comparable to that from ET.

{\bf PTA.} 
We can constrain stochastic GWs by observing residuals in arrival times of pulsar signals for a long time ($\mathcal O(10)$ years).
To put constraints on GWs, we use cross-correlations between observed pulsars.
For PTA experiments, since the inverse of the observation time is the same order of magnitude as the target frequencies, the integral in Eq.~(\ref{eq:rho_def1}) does not increase the signal-to-noise ratio much.
Therefore, it is non-trivial whether the weak-signal limit is valid or not.\footnote{ 
Note that, in Fig.~\ref{fig:gw_const_summary}, we consider $\Omega_\text{eff} h^2$ for EPTA and PTA, defined as Eq.~(\ref{eq:eff_omegagw}) in the weak-signal limit, just for comparison.
} 
Then, we redefine the signal-to-noise ratio for PTA experiments as~\cite{Anholm:2008wy,Siemens:2013zla,Chamberlin:2014ria}
\begin{align}
\rho = \sqrt{2T} &\left(\sum_{I=1}^M  \sum_{J>I}^M \chi^2_{IJ} \right)^{1/2} \nonumber \\
&\times \left[ \int^{f_\text{max}}_{f_\text{min}} \dd f \, \left(\frac{\Omega_\text{GW}(f)}{\Omega_n (f) + \Omega_\text{GW}(f)}\right)^2 \right]^{1/2},
\label{eq:pta_snr}
\end{align}
where $\chi_{IJ}$ is the Hellings and Downs coefficient for pulsars $I$ and $J$~\cite{Hellings:1983fr} (see e.g. Eq.~(13) in Ref.~\cite{Chamberlin:2014ria} for the concrete expression) and we assume that pulsars are distributed homogeneously and all pulsars have the same noise characteristics and take the average over the angle between the pulsars.\footnote{ 
Taking the average over the angle, we get
\begin{align}
\frac{1}{4\pi} \int \dd \phi \sin \theta \dd \theta  \sum_{I=1}^M  \sum_{J>I}^M \chi^2_{IJ} (\theta) = \frac{1}{48} \frac{M(M-1)}{2}. \nonumber
\end{align}
}
$\Omega_n$ is the energy density parameter for noise of each pulsar, given by
\begin{align}
\Omega_n(f) H_0^2 = \frac{2\pi^2}{3} f^3 S_n(f),
\label{eq:omega_n_def}
\end{align}
where $S_n(f)$ is the power spectral density for noise and is related to the noise power spectrum as $S_n(f) = 12\pi^2 f^2 P_n(f)$.
Here, we assume the noise power spectrum is dominated by the white timing noise as $P_n(f) \simeq 2\Delta t \sigma^2$~\cite{Thrane:2013oya}, where $1/\Delta t$ is the cadence of the measurements and $\sigma$ is the timing precision.
There have been recent observation results given by EPTA~\cite{Lentati:2015qwp}, PPTA~\cite{Shannon:2015ect}, and NANOGrav~\cite{Arzoumanian:2015liz}.
Since their constraints are comparable, we take the EPTA results as a concrete example for current constraints from PTA.
Following the result in Ref.~\cite{Lentati:2015qwp}, we take the parameters, $M=6$, $T = 18$ years, $\Delta t = 14$ days, and $\sigma = 1\mu$s for EPTA.\footnote{
Strictly speaking, since our analysis is based on the assumption that pulsars are distributed homogeneously and all pulsars have the same noise characteristics, the current constraints from EPTA in Figs.~\ref{fig:zeta_const} and \ref{fig:curv_cons_summary} are rough estimates.
However, we have checked that the constraints are almost the same as those that we derive imposing that $\Omega_\text{GW}$ for the induced GWs should not touch the constraint curve given by the black dashed line in Fig.1 in Ref.~\cite{Lentati:2015qwp}. 
}
For a future PTA project, we consider SKA as a concrete example.
We take the parameters, $M=100$, $T = 20$ years, $\Delta t = 14$ days, and $\sigma = 30$\,ns for SKA~\cite{Moore:2014lga}.

{\bf CMB, LSS, and BBN.} 
Finally, we mention the other constraints, coming from CMB, LSS, and BBN.
The constraints are based on the fact that stochastic GWs are an additional component of radiation.
The constraint from CMB and LSS is $\Omega_\text{GW} h^2 < 6.9 \times 10^{-6}$~\cite{Smith:2006nka} and from BBN is $\Omega_\text{GW} h^2 < 1.8\times 10^{-6}$~\cite{Kohri:2018awv}.
They are constraints on the total GW energy density, not the GW energy density per logarithmic interval $\Omega_\text{GW}(f)$, which means that we must compare these constraints with the induced GWs integrated over frequency, $\int_{f_\text{cut}} \, \dd \,\text{ln} f\, \Omega_\text{GW}(f)$. 
$f_\text{cut}$ is the lower cutoff of the constraint, which corresponds to $10^{-15}$ Hz for the constraint from CMB and LSS, and $10^{-10}$ Hz for that from BBN.

\section{Constraints on curvature perturbations}

\begin{figure}
\centering
\includegraphics[width=0.45 \textwidth]{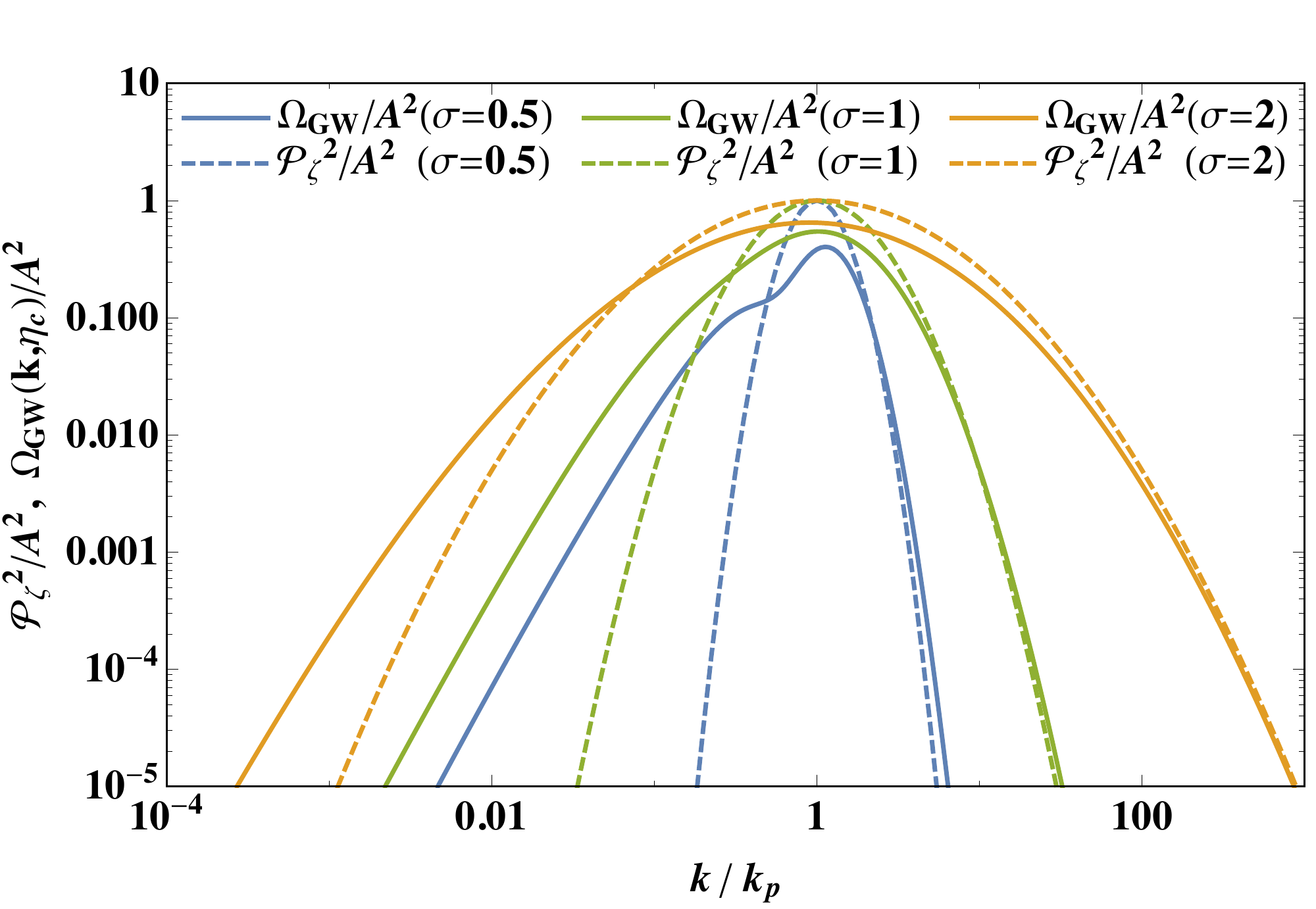}
\caption{Normalized, squared power spectra of curvature perturbations ($\mathcal P_{\zeta}^2(k)/A^2$, dashed) and induced GWs ($\Omega_\text{GW}(k,\eta_c)/A^2$, solid).
The power spectrum of the curvature perturbation is given by Eq. (\ref{eq:zeta_ansatz}) and $\sigma = 0.5$ (blue), $\sigma = 1$ (green), and $\sigma = 2$ (orange) are assumed.
Note that the quantity $\Omega_{\mathrm{GW}}(k,\eta_c)$ here does not reflect the evolution of induced GWs after their generation (see Eq. (\ref{latetime})). 
}
\label{fig:induced_gws}
\end{figure}

In this section, we explain how to derive the constraints on the curvature perturbation and present the results.
To be concrete, we parametrize the power spectrum profile of curvature perturbations as
\begin{align}
\mathcal P_{\zeta}(k) = A \exp\left( - \frac{ \left(\log k - \log k_\text{p} \right)^2}{2 \sigma^2}  \right).
\label{eq:zeta_ansatz}
\end{align}
Using Eqs.~(\ref{eq:gw_formula}) -- (\ref{eq:I_approx}), we calculate $\Omega_\text{GW}(\eta,k)$ with this spectrum.
In Fig.~\ref{fig:induced_gws}, we plot the squared power spectrum of curvature perturbations, $\mathcal P_\zeta^2$, and 
the quantity $\Omega_\text{GW}(k,\eta_c)$, both of which are normalized by $A^2$.
Here, we take $\sigma = 0.5$, $\sigma =1$, and $\sigma = 2$ as examples.
We can see that the peak height of the induced GWs, $\Omega_\text{GW}(k_p,\eta_c)$, is the same order of magnitude as $A^2$. 
This relation ($\Omega_\text{GW}(k_p,\eta_c) \sim  A^2$) is different from the previous relation used in the pioneering work ($\Omega_\text{GW}(k_p,\eta_c) \sim 30  A^2$)~\cite{Assadullahi:2009jc} by an order of magnitude.\footnote{ The work of Ref.~\cite{Assadullahi:2009jc} is based on the numerical result for a scale invariant power spectrum  ($\mathcal P_\zeta(k) = A$) in Ref.~\cite{Ananda:2006af}, which is $\Omega_\text{GW}(k,\eta_c) \simeq  33.3 A^2$. 
However, the latest result, which our work is based on, gives $\Omega_\text{GW}(k,\eta_c) \simeq  0.8222 A^2$ for the scale invariant power spectrum (see Eq.~(31) in Ref.~\cite{Kohri:2018awv}).  }
We can see that the scale dependence of $\Omega_\text{GW}(k,\eta_c)$ is very similar to that of ${\cal P}_\zeta^2$ on the scales smaller than the peak scale ($k_p^{-1}$).
Meanwhile, on larger scales, the GWs decay as $\Omega_\text{GW} h^2 \propto k^{3}$ even though the curvature power spectrum decays more rapidly.

We derive (expected) limits on $A$ for each $\sigma$ and $k_p$ by finding the value of $A$ which yields the the signal-to-noise ratio $\rho$, given by Eq.~(\ref{eq:rho_def}) for interferometer experiments or Eq.~(\ref{eq:pta_snr}) for PTA observations, unity, taking into account $\Omega_\text{GW,eff}$ for each observation discussed above, except for the CMB and BBN constraints.
We take $T=18$ years for EPTA, $T=20$ years for SKA, $T=4$ months for aLIGO (O2)~\cite{Abbott:2017xzg}, and $T=1$ year for the others as fiducial values. 
For CMB and BBN constraints, we derive the limits by finding the value of $A$ which makes the integral $\int \dd \, \text{ln} f \, \Omega_\text{GW}(f)$ equal to the $\Omega_\text{GW}$ constraints, plotted in Fig.~\ref{fig:gw_const_summary}.

\begin{figure*}[htbp]
  \begin{center}
    \begin{tabular}{c}

      \begin{minipage}{0.45\textwidth}
        \begin{center}
          \includegraphics[width=\hsize]{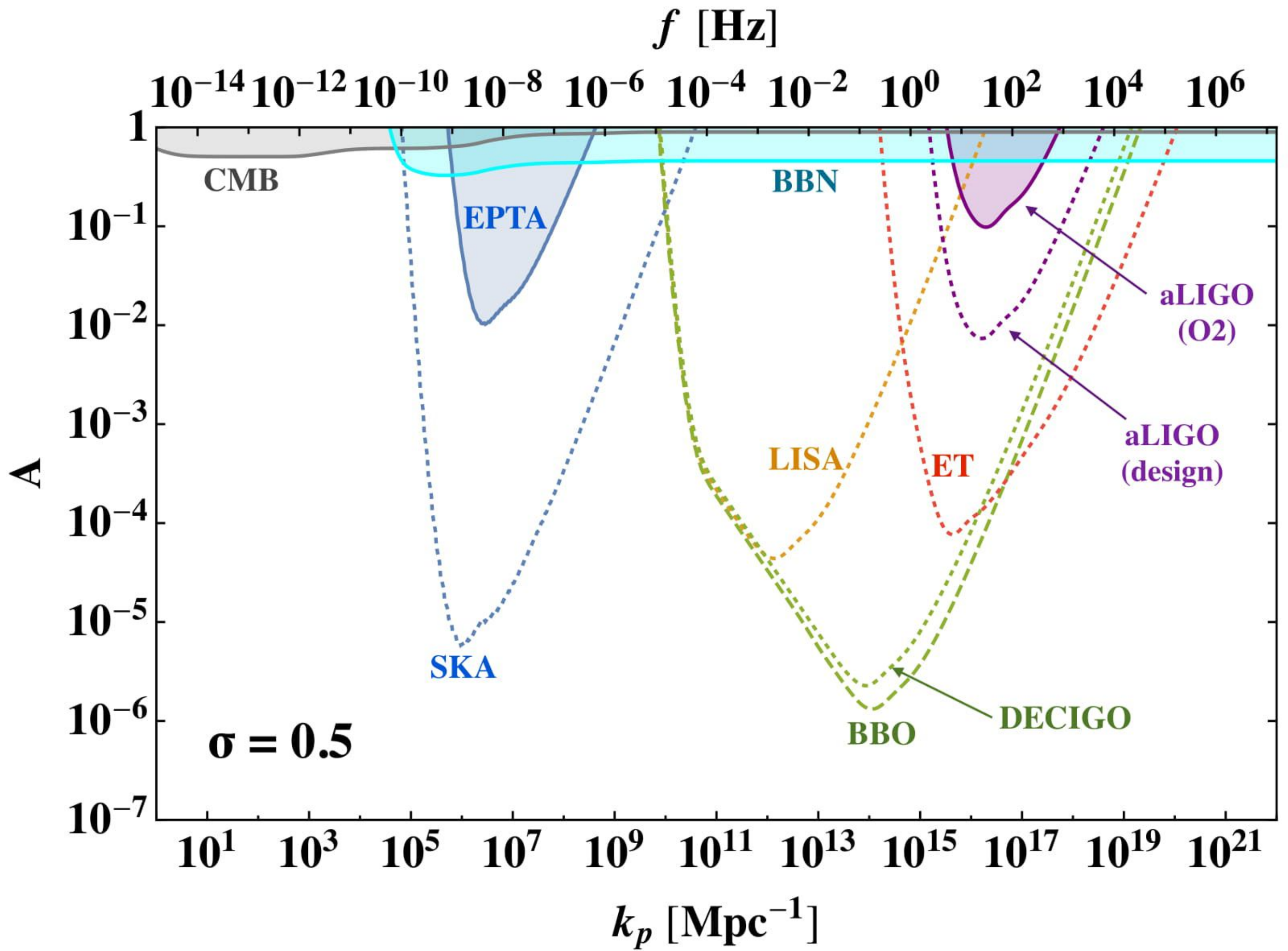}
        \end{center}
      \end{minipage}
	\hspace{0.5cm} 
      \begin{minipage}{0.45\textwidth}
        \begin{center}
          \includegraphics[width=\hsize]{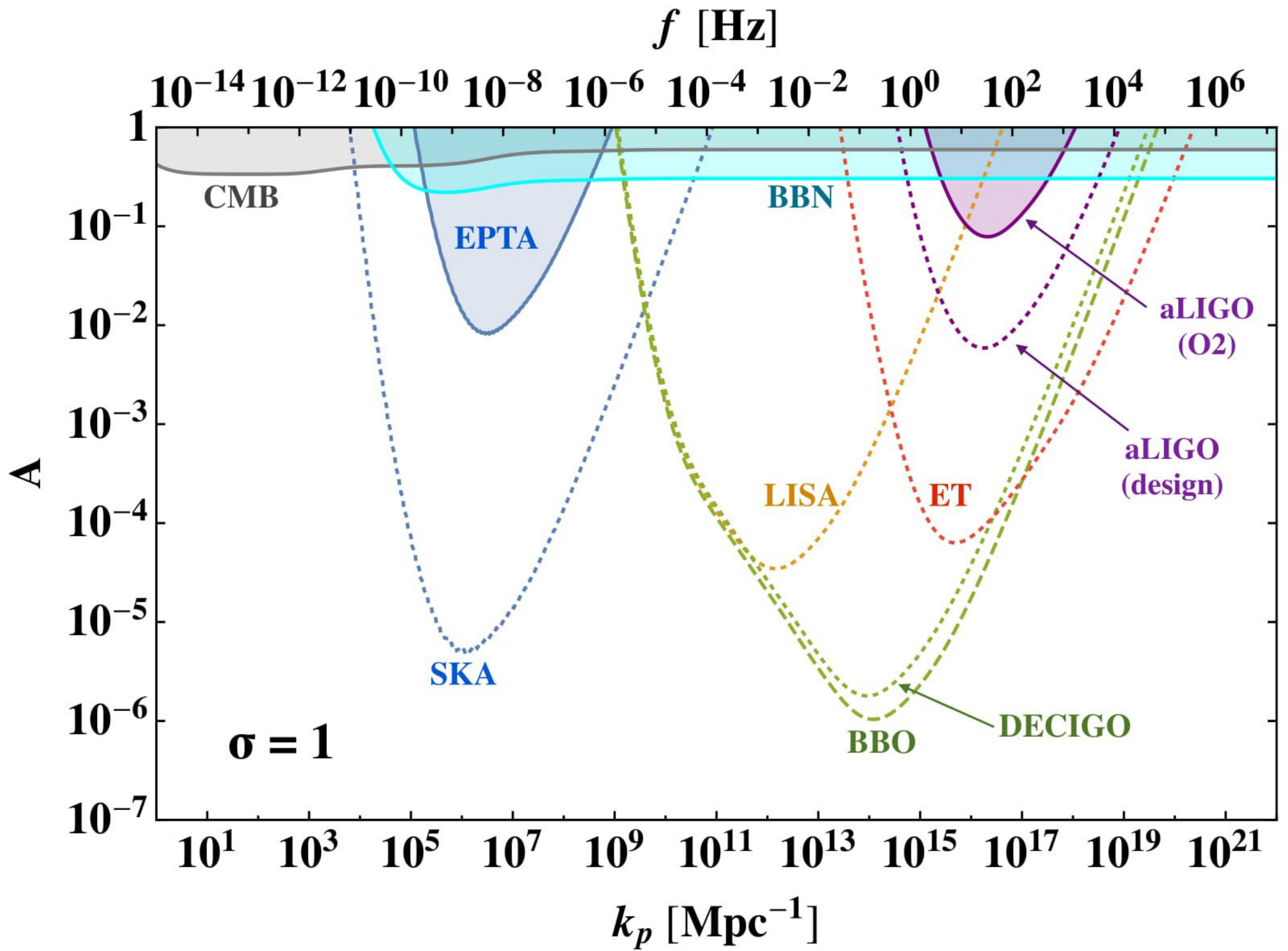}
        \end{center}
      \end{minipage}\\
      
      \begin{minipage}{0.45\textwidth}
      \vspace{0.6cm}
        \begin{center}
          \includegraphics[width=\hsize]{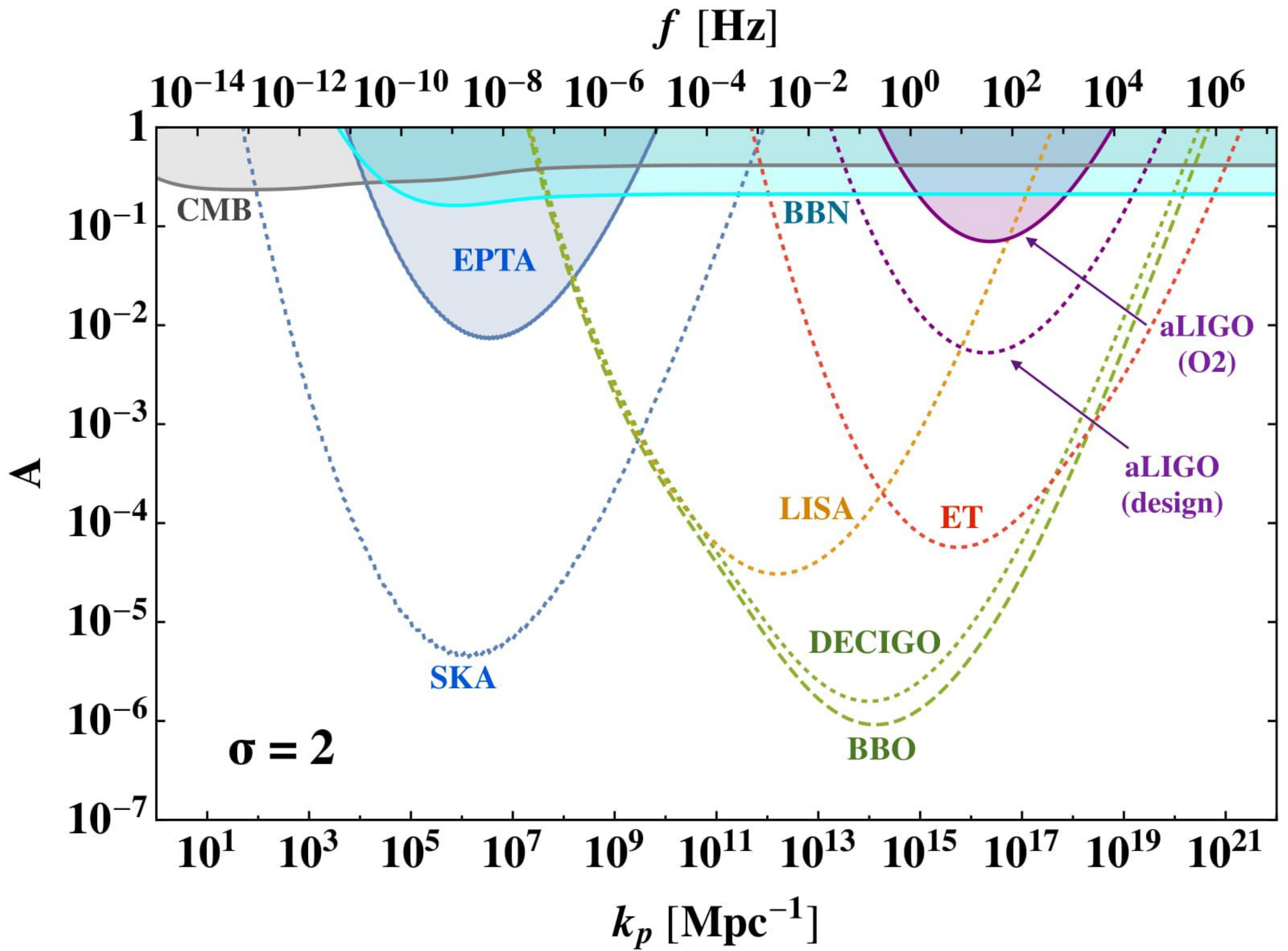}
	        \end{center}
      \end{minipage}
    \end{tabular}
    \caption{\small	
Limits on curvature perturbations with $\sigma = 0.5$ (upper left figure), $\sigma = 1$ (upper right figure) and $\sigma = 2$ (lower figure). 
The vertical axis is $A$ and the lower horizontal axis is $k_p$, which are defined in Eq.~(\ref{eq:zeta_ansatz}).
The upper horizontal axis shows the frequency corresponding to $k_p$.
The colors and styles of the curves here correspond to those in Fig.~\ref{fig:gw_const_summary} (e.g. the blue solid lines show the constraints from the current PTA observations).
The shaded regions are excluded by the current observations, as in Fig.~\ref{fig:gw_const_summary}.
	}	
	\label{fig:zeta_const}
  \end{center}
\end{figure*}

Figure~\ref{fig:zeta_const} shows the limits on $A$ for $\sigma = 0.5,1$ and 2.\footnote{ When we obtain the plots in Fig.~\ref{fig:zeta_const}, we find that, for SKA curves, both results based on Eqs.~(\ref{eq:rho_def}) and (\ref{eq:pta_snr}) are almost the same, which means the weak-signal limit is a good approximation. This is mainly because the large number of pulsars increases the signal-to-noise ratio sufficiently so that $\rho=1$ is reached in the weak-signal regime.} 
The parameter space of the primordial spectrum that can be constrained by GW experiments is wider when $\sigma$ is larger, due to the spread of the GW spectrum as shown in Fig.~\ref{fig:induced_gws}. The shaded regions show the constraints from the existing data of current observations.
In particular, the current PTA observations constrain the perturbations as $A\lesssim \mathcal O(10^{-2})$ on $k \sim \mathcal O(10^6)$Mpc$^{-1}$.
The noticeable scale dependence of constraints from CMB and BBN is due to the change in the relativistic degrees of freedom and the frequency cutoff of the constraints. 
As to future prospects, the amplitude of the curvature perturbations could be investigated over a wide range of scales.
In particular, the curvature perturbations with $\mathcal P_\zeta =\mathcal O(10^{-4}) - \mathcal O(10^{-6})$ could be observed or constrained by SKA, LISA, BBO, or ET.
Note that, although we assume the concrete observation times and signal-to-noise ratio to derive Fig.~\ref{fig:zeta_const},
 the parameter dependence of the constraints in the weak-signal limit is given by $A\propto \rho^{1/2} T^{-1/4}$, which we can easily see from Eq.~(\ref{eq:rho_def}).

\section{Discussion and Conclusion}

In this paper, we have revisited the constraints on curvature perturbations from the GWs induced at second order in curvature perturbations.
If the curvature perturbations are large enough, GWs induced from the second order perturbations could exceed the existing or future limits on GWs, which means the curvature perturbation can be constrained. By using existing data and sensitivity curves for future experiments, we have derived the existing and expected limits on the curvature perturbation. 

The constraints from the induced GWs were also discussed in Ref.~\cite{Assadullahi:2009jc}.
Our updates are roughly divided into two aspects.
First, we have used updated equations for the induced GWs, and in addition made the discussions more precise, by numerically calculating the spectrum of induced GWs, assuming a concrete shape of the primordial power spectrum. 
The induced GWs predicted by these updated equations differ from those predicted by the equation used in the previous study by an order of magnitude.
Second, we have performed more precise analysis when we constrain the induced GWs from the sensitivity curves of experiments.
To take into account the frequency dependence of the induced GWs and sensitivity curves correctly, we have calculated the signal-to-noise ratio, defined in Eq.~(\ref{eq:rho_def}) or Eq.~(\ref{eq:pta_snr}).
We have also discussed the dependence of the limits on the shape of the power spectrum.

In this work, we have simply assumed the null detection of GWs for each experiment, but in future stochastic GWs of astrophysical origins, or stochastic GWs of cosmological origins that are different from the induced GWs we have considered, such as those from quantum vacuum fluctuations during inflation, first-order phase transitions, or cosmic strings~\cite{Caprini:2018mtu}, may be detected. In such a case, the limits on the curvature perturbation would be affected, and a discussion about this issue depends on the experiment. For instance, at relatively low frequencies, relevant to SKA, stochastic GWs from mergers of supermassive black holes would be important. An estimation of such GWs inevitably involves uncertainties stemming from complex astrophysical processes, but the amplitude of $\Omega_{\mathrm{GW}}h^2\sim 10^{-11}(f/10^{-8}\mathrm{Hz})^{2/3}$  was noted to be a conservative lower limit \cite{Caprini:2018mtu}, based on Refs. \cite{Sesana:2008mz,Sesana:2012ak}. If stochastic GWs from supermassive-black-hole mergers are indeed detected, stochastic GWs of cosmological origins, including induced GWs, would be buried, and this implies less information obtained about the early Universe. For instance, the limits on the curvature perturbation based on induced GWs would be weaker than those obtained from the null detection of stochastic GWs. Naively in this case one may constrain the curvature perturbation by requiring induced GWs to be less than the detected GW background from supermassive-black-hole-binary mergers. We may do a bit better than that by making use of the anisotropy of stochastic GWs from supermassive black holes, which is at level of $\sim 20\%$ of the isotropic component \cite{Mingarelli:2017fbe,Barack:2018yly,Taylor:2013esa,Taylor:2015udp}. One might also be able to improve the limits by making use of characteristic non-Gaussianity of induced GWs \cite{Bartolo:2018evs,Bartolo:2018rku}.  As another example, in the case of BBO, it may be possible to detect and subtract out $\sim 3\times 10^5$ merging binaries composed of neutron stars and/or black holes, out to $z\sim 5$ \cite{Cutler:2009qv}. If cosmological, stochastic GWs are indeed detected as a result of successful subtraction of astrophysical foregrounds, then differentiating between different kinds of cosmological GWs  using their properties such as the spectrum, non-Gaussianity and chirality \cite{Caprini:2018mtu},  would be crucial, and one of GW origins here is the induced GWs we have discussed. In this case, if we fail to identify the source of the detected cosmological GW background, one would obtain the limit on the curvature perturbation by simply requiring induced GWs to be less than the detected GWs. Instead if we can reliably exclude the possibility that the detected GWs are induced GWs, one may obtain limits tighter than that, possibly making use of non-Gaussianity, and finally if we can conclude that the detected GWs \textit{are} induced GWs, we would be able to determine the power spectrum of curvature perturbation. See also Ref.~\cite{Regimbau:2016ike} for subtraction of astrophysical foregrounds to detect cosmological GWs by ground-based detectors.

\begin{figure}[h]
\centering
\includegraphics[width=0.45 \textwidth]{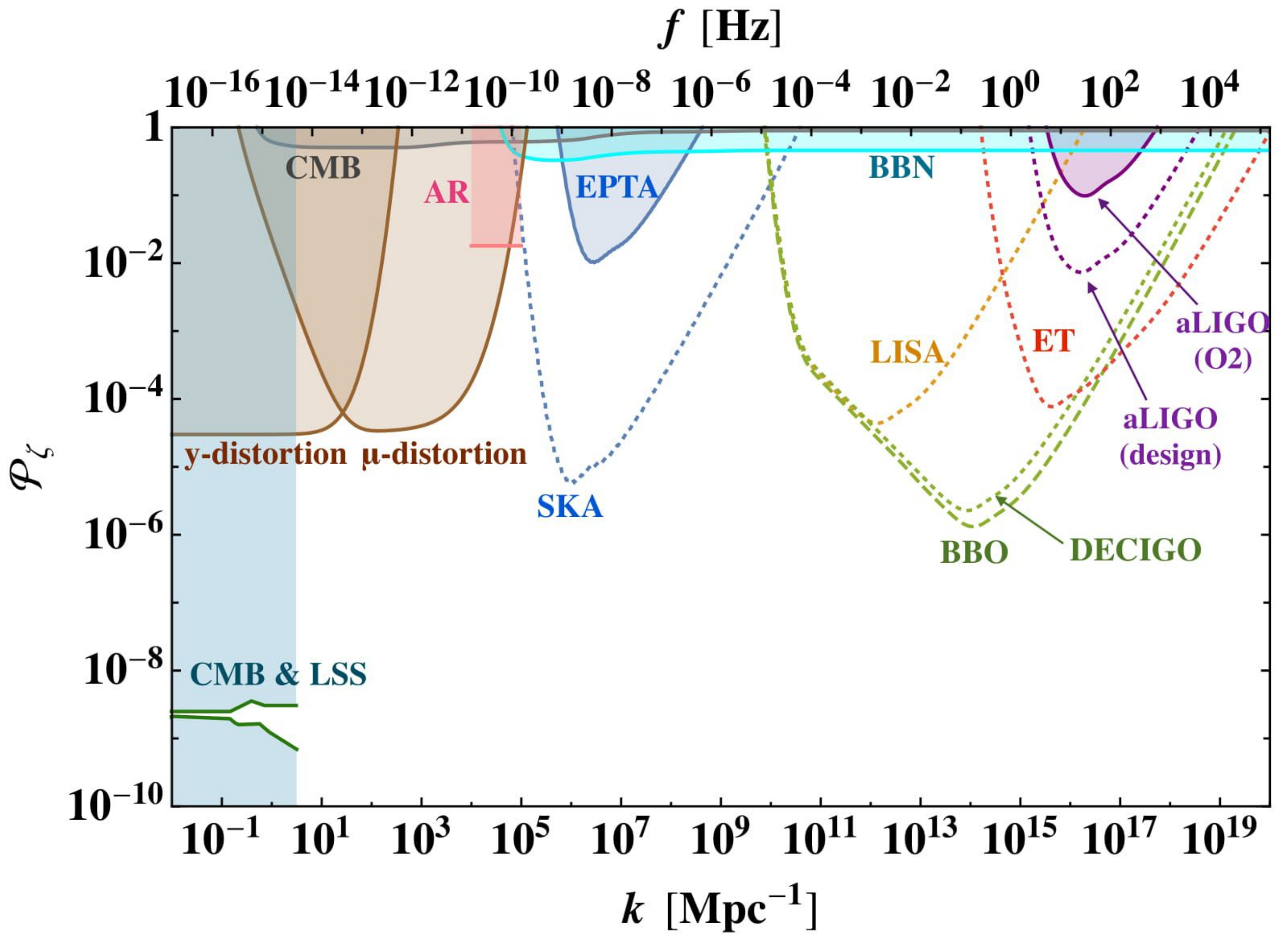}
\caption{Existing and expected limits on the small-scale power spectrum of the curvature perturbation.
The constraints from the induced GWs are the same as those shown in Fig.~\ref{fig:zeta_const} ($\sigma=0.5$). 
In addition to the constraints derived in this paper,
the constraints from acoustic reheating (AR)~\cite{Inomata:2016uip} (pink, see also~\cite{Jeong:2014gna} and~\cite{Nakama:2014vla}), CMB spectral distortions~\cite{Chluba:2012we,Kohri:2014lza} (brown), and CMB/LSS observations~\cite{Hunt:2015iua} (dark green) are also plotted. 
The shaded regions are excluded by the current observations, whereas expected limits from future experiments are shown by the dashed and dotted lines.
}
\label{fig:curv_cons_summary}
\end{figure}

Before closing, let us summarize the current status and future prospects for probing small-scale primordial perturbations.
Figure~\ref{fig:curv_cons_summary} shows the current and future expected limits on the small-scale curvature perturbations.
In this figure, for the constraints from the induced GWs, the vertical and the horizontal axis should be understood as $A$ and $k_p$, which are the amplitude and position of a sharp spike of the power spectrum of the curvature perturbation we have used in this paper. 
This makes a simple comparison of different limits possible, which would be instructive. 
We take $\sigma=0.5$ to show the conservative constraints.
In addition to the constraints by the induced GWs, we also plot the conventional constraints from CMB and LSS observations, CMB distortions, and acoustic reheating.
We derive the constraints from CMB distortions performing the integration of Eq.~(10) in Ref.~\cite{Chluba:2012we} with the profile of the power spectrum given in Eq.~(\ref{eq:zeta_ansatz}) in this paper, using the limits on the $\mu$ and $y$ parameters obtained by COBE/FIRAS, which are $\mu\lesssim9\times 10^{-5}$ and $y\lesssim1.5\times 10^{-5}$ \cite{Fixsen:1996nj}. 
We do not show constraints from UCMHs and PBHs because the constraints from such objects have some uncertainties, as we mentioned in Sec.~\ref{sec:intro}.
We close by concluding that the future GW observations will shed new light on the small-scale primordial spectrum, which can not be probed by other observations.

\acknowledgments 

We are grateful to Jun'ichi Yokoyama, Sachiko Kuroyanagi, and Christian Byrnes for helpful communications. We thank Johns Hopkins University and Research Center for the Early Universe, University of Tokyo, for hospitality received during this work. 
This work was supported by World Premier International Research Center Initiative (WPI Initiative), MEXT, Japan.
K.I. is supported by 
Advanced Leading Graduate Course for Photon Science, 
and JSPS Research Fellowship for Young Scientists.

\begin{center}
{\bf Note Added }
\end{center}

A related paper~\cite{Byrnes:2018txb} appeared when we were finalizing our work. 
Though their primary focus is PBHs, figures showing limits on the primordial spectrum including those from the induced GWs are also shown. They included only current and future PTA and LISA and also did not perform the analysis that we have done in this paper, calculating the signal-to-noise ratio. 


\appendix
\section{Relation between $k$ and $T$}
\label{app:relation_k_t}
In this appendix, we present a derivation of Eq.~(\ref{eq:relation_k_t}). 
The Friedmann equation is given by 
\begin{align}
	3M_\text{Pl}^2 H^2 = \rho.
\end{align}
For the left-hand side, we will shortly use $a H \simeq 1/\eta$, satisfied during the RD era, and 
for the right hand side, we will use $\rho_r \propto gT_\gamma^4$.
On the other hand, around the matter-radiation equality, the scale factor is given by~\cite{Mukhanov:991646}
\begin{align}
	a(\eta) = a_\text{eq} \left( \left( \frac{\eta}{\eta_*} \right)^2 + 2 \left( \frac{\eta}{\eta_*} \right) \right),
\end{align}
where $\eta_* = \eta_\text{eq}/(\sqrt{2} -1 )$.
Then we find $a_\text{eq} H_\text{eq} = 2 (2-\sqrt{2})/\eta_\text{eq}$.
Note also the relations $\rho_\text{eq} = 2\rho_{r,\text{eq}}$ and $\rho_{\gamma,\text{eq}} \propto g_\text{eq} T_\text{eq}^4$.
First, from the Friedmann equation, 
\begin{equation}
    \frac{a H}{a_\text{eq} H_\text{eq}} = \frac{a}{a_\text{eq}} \sqrt{ \frac{\rho}{2 \rho_{\gamma,\text{eq}}}},
\end{equation}
which can be rewritten as 
\begin{equation}
    \frac{1}{2(2-\sqrt{2})} \frac{\eta_\text{eq}}{\eta} =  \frac{1}{\sqrt{2}} \left(\frac{g_\text{s,eq}}{g_\text{s}}\right)^{1/3} \left( \frac{g}{g_\text{eq}} \right)^{1/2} \frac{T}{T_\text{eq}},
\end{equation}
where we have used the entropy conservation relation: $g_{s,\text{eq}} a^3_{\text{eq}} T^3_\text{eq} = g_s a^3 T^3$. Then finally we find that the comoving wavenumber which reenters the horizon at $\eta$, that is,  $k\eta=1$, is related to the temperature at that moment via
\begin{equation}
    \frac{k}{k_\text{eq}} = 2(\sqrt{2} -1)  \left(\frac{g_\text{s,eq}}{g_\text{s}}\right)^{1/3} \left( \frac{g}{g_\text{eq}} \right)^{1/2} \frac{T}{T_\text{eq}}.
\end{equation}
Note again that this relation is valid during the RD era.

\small
\bibliographystyle{apsrev4-1}
%

\end{document}